\documentclass[
reprint,
amsmath,amssymb,
]{revtex4-2}

\usepackage{graphicx}
\usepackage{dcolumn}
\usepackage{bm}
\usepackage{physics}
\usepackage{braket}
\usepackage{qcircuit}
\usepackage{algorithmic}
\usepackage{algorithm}
\usepackage{overpic}

\newcommand{\cH}{\mathcal{H}}
\DeclareMathOperator*{\argmax}{arg\,max}
\DeclareMathOperator{\sgn}{sgn}

\begin{document}

\title{Enhancing Quantum Annealing in Digital-Analog Quantum Computing} 

\author{Tadashi Kadowaki}
\affiliation{Global R\&D Center for Business by Quantum-AI Technology (G-QuAT), National Institute of Advanced Industrial Science and Technology (AIST), 1-1-1 Umezono, Tsukuba, Ibaraki 305-8568, Japan.}
\email{tadashi.kadowaki@aist.go.jp}
\affiliation{DENSO CORPORATION, 1-1-4, Haneda Airport, Ota-ku, Tokyo 144-0041, Japan}

\date{\today}

\begin{abstract}
Digital-analog quantum computing (DAQC) offers a promising approach to addressing the challenges of building a practical quantum computer. By efficiently allocating resources between digital and analog quantum circuits, DAQC paves the way for achieving optimal performance. We propose an algorithm designed to enhance the performance of quantum annealing. This method employs a quantum gate to estimate the goodness of the final annealing state and find the ground state of combinatorial optimization problems. We explore two strategies for integrating the quantum annealing circuit into the DAQC framework: (1) for state preparation, and (2) for embedding within the quantum gate. While the former strategy does not yield performance improvements, we discover that the latter enhances performance within a specific range of annealing time. Algorithms demonstrating enhanced performance utilize the imaginary part of the inner product of two states from different quantum annealing settings. This measure reflects not only the energy of the classical cost function but also the trajectory of the quantum dynamics. This study provides an example of how processing quantum data using a quantum circuit can outperform classical data processing, which discards quantum information.
\end{abstract}

\maketitle

\section{Introduction}
\label{sec1}

Fault-tolerant quantum computing demands vast quantum and classical resources, such as physical qubits consisting of a logical qubit, fast and large analog and digital circuits, and registers for error collection. Digital-analog quantum computing (DAQC)~\cite{Dodd2002} provides an efficient avenue for resource utilization in the noisy intermediate-scale quantum era, thereby accelerating the application of quantum computations like quantum chemistry~\cite{Yung2014}, Fourier transformation~\cite{Martin2020}, optimization~\cite{Headley2022}, and quantum simulation~\cite{Celeri2021,Guseynov2022}. By integrating qubit rotation and two-qubit Hamiltonian, e.g., homogeneous Ising Hamiltonian, DAQC enables universal computation where the two-qubit dynamics are executed as an analog computation~\cite{Parra-Rodriguez2020}. Studies into noise tolerance~\cite{Garcia-Molina2021,Celeri2021} and hardware implementation~\cite{Mezzacapo2014,Lamata2018,Galicia2020,Gonzalez-Raya2021,Yu2022,Kumar2023} have been conducted, and a recent study illustrates a quantum approximate optimization algorithm within the DAQC framework~\cite{Headley2022}. Despite these advancements, the potential of the framework remains to be fully explored.

In this study, we investigate quantum annealing (QA) within a DAQC framework to enhance the probability of finding the ground state of a problem Hamiltonian. Given the recent advancements in the quantum gate and quantum annealer technologies, our proposed circuit has practical implementation potential. By investigating QA within the DAQC framework, we contribute new insights that will expand the DAQC framework in quantum computing and pave the way for further research in this field.

QA is a quantum algorithm and a well-studied component of analog quantum computing~\cite{Kadowaki1998, Kadowaki1998a, Brooke1999, Santoro2002, Santoro2006, Das2008, Morita2008, Tanaka2017, Farhi2001}. It is inspired by simulated annealing (SA)~\cite{Kirkpatrick1983}, which explores solution space by applying scheduled thermal fluctuations that decrease from high to low temperature. Initially, SA algorithm searches the solution space globally and randomly at high temperatures, then locally at low temperatures. If the cooling rate is appropriately slow, SA can find the global minimum solution. In QA, quantum fluctuations replace thermal fluctuations. These fluctuations are controlled by transitioning the system’s Hamiltonian from a driver Hamiltonian to a problem Hamiltonian, which serves as the cost function of the problem. Conventional QA employs a transverse field for the driver Hamiltonian, as the ground state is a superposition of 0 and 1 for all $N$-qubits, represented as $\ket{+}^{\otimes N}$. According to the adiabatic theorem, if the Hamiltonian's transition schedule is sufficiently slow, we will ultimately obtain the ground state, or the exact solution, of the cost function.

A significant drawback of SA and QA is the lengthening of annealing time required to obtain the exact solution for complex instances. In these cases, the energy gap between the ground and the excited states diminishes exponentially as a function of system size, thereby requiring an exponential increase in annealing time as per the adiabatic theorem. When we exceed the adiabatic condition by accelerating the schedule, the quantum system deviates from the adiabatic evolution path, which reduces the probability of finding the ground state. This issue can be mitigated through counter-diabatic (CD) driving~\cite{Demirplak2003, Demirplak2005, Berry2009, Chen2010, Chen2011, Takahashi2013, Jarzynski2013a, Torrontegui2012, Campo2013, Takahashi2017, Sels2017, Ozguler2018, Claeys2019, Hartmann2019, Guery2019, Prielinger2021, Hatomura2021, Imoto2021}. An additional CD Hamiltonian provides an extra force that guides the system along the adiabatic path of the original QA Hamiltonian, even under accelerated conditions. To construct the exact form of the CD driving Hamiltonian, theoretically, all eigenstates of the QA Hamiltonian are needed. However, due to the computational complexity of this task, an approximate Hamiltonian, such as a local $y$-field with suitable time-dependent coefficients, was proposed instead~\cite{Sels2017}. This approach enabled the local $y$-field to be optimized through a variational method.

Quantum Greedy Optimization (QGO) is an algorithm, which extends QA by the variational CD driving Hamiltonian~\cite{Kadowaki2023}. This algorithm provides a method to determine the variational CD parameters by conducting a series of sensitivity analyses. A simplification of this variational CD Hamiltonian is characterized by a time-dependent and site-dependent $y$-field with variational parameters. In this method, the initial state in the $x$-direction, a superposition of 0 and 1, is deliberately rotated towards the $\pm z$-direction by independently applying the $\mp y$-field at each site. The energy can be reduced through the appropriate selection of the $y$-field direction. As a result, the signs of the parameters are intrinsically linked to the ground state. The algorithm sequentially optimizes these parameters through iterative energy sensitivity analysis for each parameter.

In this paper, we examine QGO, an extension of the QA algorithm, within the DAQC framework. Henceforth, we refer to this algorithm as Digital-analog Quantum Greedy Optimization (DAQGO). While energy is primarily used as a measure in the sensitivity analysis of QGO, other observable measures of performance are also applicable. For instance, fidelity was tested in a previous study~\cite{Kadowaki2023}, and it demonstrated greater performance than energy. Fidelity, which represents the overlap between the final state of QA and the ground state in optimization settings, ensures that QGO consistently finds the ground state. Although this measure is unrealistic, it indicates that an effective measure can indeed outperform the energy measure. Therefore, our evaluation criteria are expanded to include a variety of metrics from quantum gate circuits, beyond just energy considerations. Specifically, during our sensitivity analysis on the parameter, we focus on the phase difference. This metric is derived by calculating the inner product between the final states from two QA outputs with different parameter settings.

We investigate two strategies for integrating the QA circuit into the DAQC framework: (1) using it for state preparation and (2) embedding it within the quantum gate circuit. For the first strategy, we construct a gate circuit to evaluate the approximated energy difference between two states. The second strategy involves testing two types of circuits, one for determining the energy difference and the other for calculating the inner product of the two states. Our findings suggest that while the circuits for energy difference do not enhance performance, the circuit for the inner product does improve performance within a certain range of annealing time. Furthermore, we calculate the number of required shots for different instances and conduct experiments with 2-qubit problems on a trapped ion quantum device.

Through this paper, we demonstrate the extended QA functionality. While our focus is on analog device implementation, this does not limit the application of the functionality exclusively to such devices. Depending on technological advancements, digital versions of shortcuts to adiabaticity can also be utilized~\cite{Barends2016,Hegade2021}.

This paper is organized as follows. In Section \ref{sec2}, we describe the algorithms using QA within the DAQC framework. In Section \ref{sec3}, we compare the performance of these algorithms with that of vanilla QGO and traditional QA. This section also includes results from experiments conducted on the trapped ion quantum device. The final section provides a discussion and analysis of our results.

\section{Quantum Greedy Optimization and Digital-Analog Quantum Circuits}
\label{sec2}

In this section, we present DAQC circuits that combine QA with digital quantum gates for optimization problems. Two extensions are proposed for the QA in the DAQC circuit: (a) the use of the $y$-field as a CD driving source, and (b) functioning as a controlled unitary. The former was introduced in previous work, while the latter is a novel extension with two inputs: a control port and an initial state port for QA. For simplicity, we refer to both conventional and extended QAs simply as QA without further distinction. The QGO algorithm utilizes measures from the proposed DAQC circuits. In the subsequent subsections, we will review the QGO algorithm, outline the DAQC circuits used for energy and other measures, and describe the experimental setups.

\subsection{QGO algorithm}

In strategies for fixing variables in optimization problems, such as roof duality~\cite{Kolmogorov2004, Boros2002, Boros2006, Rother2007}, sample persistency~\cite{Karimi2017}, and hybrid quantum annealing~\cite{Irie2021}, each variable is prioritized based on the difficulty of calculating its optimal value. That is, some variables can be determined individually without considering other variables. Some variables depend heavily on a large set of other variables due to their strong interactions. Therefore, an extensive search in the solution sub-space covered by these dependent variables is required to calculate their optimal values. The rest depend on a smaller set of variables and require an extensive search in the smaller sub-space of these variables. Sequential optimization in QGO applies this assumption repeatedly, identifying a variable to be fixed and determining its optimal value until all variables are fixed.

The total Hamiltonian of QGO is given by:
\begin{equation}
    \cH = A(t) \cH^z + B(t) \cH^x + \sum_i C_i(t) \cH_i^y ,
\end{equation}
where $\cH^z$ is the problem Hamiltonian, $\cH^x$ is the transverse field, and $\cH_i^y$ is the CD driving Hamiltonian. Together, $\cH^z$ and $\cH^x$ form the QA Hamiltonian. These Hamiltonians are defined as follows:
\begin{gather}
    \cH^z = - \sum_{i<j} J_{ij} \sigma_i^z \sigma_j^z - \sum_i h_i \sigma_i^z , \\
    \cH^x = - \sum_i \sigma_i^x , \\
    \cH_i^y = - \sigma_i^y .
\end{gather}
In these equations, $\sigma_i^k (k \in { x, y, z})$ denotes a Pauli matrix. The scheduling parameters $A(t)$ and $B(t)$ are functions of time, given by $A(t) = at/\tau$ and $B(t) = b(1-t/\tau)$, where $a$ and $b$ are parameters, and $\tau$ represents the annealing time. For the sake of simplicity, we set $a = 1$ in the following discussions. The variational parameters $\{c_i\}$ characterize the $y$-field, defining $C_i(t) = c_i \sin^2 (\pi t/\tau)$.

The initial state of the system aligns with the $x$-direction, matching the ground state of the initial Hamiltonian, $\cH^x$. Depending on the sign of the parameter $c_i$, the $y$-field rotates this initial state towards either the $+z$ or $-z$ direction, both of which are in classical solution space. Given the appropriate parameters $\{c_i\}$, each qubit rotates towards the correct direction of the ground state in a short annealing time scenario (the CD regime). Hence, finding the correct parameters is equivalent to finding the ground state.

The QGO algorithm identifies a variable to fix from sensitivity analyses of the unfixed variables. The absolute value of the sensitivity is used for variable selection, and we select the most sensitive variable to be fixed. Then, the sign and absolute value of the parameter $c_i$ are set by adopting the sign of the sensitivity and a pre-calculated value, $c_{\mathrm{opt}}$, respectively. As we evaluate sensitivities for $N+1-i$ parameters in the $i$-th iteration, the total number of evaluations is $\sum_{i=1}^{N-1} (N+1-i) = (N+2)(N-1)/2$, where $N$ represents the number of variables in the optimization problem. The value of $c_{\mathrm{opt}}$ is approximated using $c_{\mathrm{opt}}^N$, which represents the optimal value derived from a ferromagnetic model with system size $N$. Similarly, the optimal value of $b_{\mathrm{opt}}$ is approximated by $b_{\mathrm{opt}}^N$. The detailed procedure is outlined in Algorithm~\ref{alg1}.

Two measures, energy and fidelity, have been studied. In the context of optimization problems, fidelity refers to the overlap between the final state and the ground state. While this measure may not be practical for real-world applications, QGO consistently identifies the ground state using this measure. These results suggest that there might exist a measure that exhibits better performance than energy alone. In this section, we will introduce candidate measures. Since sensitivity analysis is calculated based on the finite difference of energy or other measures, the following subsections will explain how to calculate them using quantum circuits. 

\begin{algorithm}[H]
    \caption{QGO algorithm}
    \label{alg1}
    \begin{algorithmic}[1]
    \REQUIRE number of variables $N$, goodness measure $f(b, {\bm c})$, differentiation interval $h$, optimal values from ferromagnetic system $b_{\rm opt}^N$ and $c_{\rm opt}^N$
    \ENSURE a solution of the cost function
    \STATE $b \leftarrow b_{\rm opt}^N$
    \STATE ${\bm c} \leftarrow (0, \cdots, 0)$
    \REPEAT
    \STATE $g_i \leftarrow f(b, c_1, \cdots, c_i+h, \cdots, c_N) - f(b, {\bm c})$
    \STATE $k \leftarrow \argmax_{j \in \{j|c_j=0\}} | g_j |$
    \STATE $c_k \leftarrow - c_{\rm opt}^N \sgn g_k$
    \UNTIL $c_i \ne 0$ for all $i$
    \RETURN $\sgn {\bm c}$
    \end{algorithmic}
\end{algorithm}

\subsection{DAQGO1: QA as a state preparation}

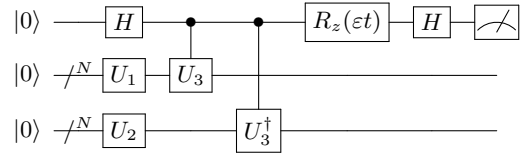
\begin{figure}[h]
    \centering
    \leavevmode
    \Qcircuit @C=1em @R=.7em {
      \lstick{\ket{0}} & \qw & \gate{H} & \ctrl{1} & \ctrl{2} & \gate{R_z(\varepsilon t)} & \gate{H} & \meter \\
      \lstick{\ket{0}} & {/^N} \qw & \gate{U_1} & \gate{U_3} & \qw & \qw & \qw & \qw \\
      \lstick{\ket{0}} & {/^N} \qw & \gate{U_2} & \qw & \gate{U_3^\dag} & \qw & \qw & \qw \\
    }
    \caption{Quantum circuit for DAQGO1. $U_1$ and $U_2$ are QA blocks used for the preparation of states with and without the $y$-field. $U_3$ and $U_3^\dag$ are the controlled Hamiltonian simulation blocks. The bottom 2 lines consist of $N$ qubits.}
    \label{fig1}
\end{figure}

A simple quantum circuit for calculating the energy difference between the test and reference states is depicted in Fig.~\ref{fig1}. In line 4 of Algorithm~\ref{alg1}, we calculate $g_i$ by testing all unfixed variables $\{c_i|c_i=0\}$. Hereafter, we provide a simplified description of how the energy difference or other measures are calculated for a target variable. A superposition of $\ket{0}$ and $\ket{1}$ on an ancilla qubit is generated by a Hadamard gate,
\begin{equation}
    H = \frac{1}{\sqrt{2}} \begin{pmatrix} 1 & 1 \\ 1 & -1 \end{pmatrix} .
\end{equation}
The $U_1$ and $U_2$ blocks perform quantum evolution given by $T \int_0^\tau e^{-i \cH t}$, where $T$ denotes the time-ordered product, with and without the $y$-field to prepare the test and reference states $\ket{\psi_T} = U_1 \ket{0} = \sum_i \alpha_i \ket{u_i}$ and $\ket{\psi_R} = U_2 \ket{0} = \sum_j \beta_j \ket{u_j}$, respectively. Both blocks use the same annealing time $\tau$. The Hamiltonian simulation blocks $U_3 = \exp(-i\cH^z t)$ and $U_3^\dag = \exp(+i\cH^z t)$, controlled by the ancilla qubit, evolve the states according to the problem Hamiltonian $\cH^z$. The controlled gates of $U$ with a closed and open circles are represented as
\begin{equation}
    \ket{0} \bra{0} \otimes I + \ket{1} \bra{1} \otimes U ,
    \label{eq6}
\end{equation}
and
\begin{equation}
    \ket{0} \bra{0} \otimes U + \ket{1} \bra{1} \otimes I ,
    \label{eq7}
\end{equation}
respectively, where the matrices are divided into sub-spaces by the ancilla qubit. An additional rotation gate,
\begin{equation}
    R_z(\varepsilon t) = \begin{pmatrix} 1 & 0 \\ 0 & e^{i\varepsilon t} \end{pmatrix} ,
\end{equation}
shifts the phase of the $\ket{1}$ state in the ancilla qubit, where $0 \leq \varepsilon \ll 1$.

The algorithm employs QAs for $U_1$ and $U_2$, which have been enhanced with the integration of the $y$-field as a CD term, as described in~\cite{Kadowaki2023}. In contrast, $U_3$ is developed within a quantum circuit. The Hamiltonian for $U_3$ is composed entirely of one-body and two-body $\sigma^z$ terms, enabling efficient implementation of the circuit with a depth of $O(N^2)$.

Before measurement, an Hadamard gate is applied. The final state is given by
\begin{eqnarray}
    \ket{\psi}_F & = & \frac{1}{2} \ket{0} \left( \ket{\psi_T} \ket{\psi_R} + e^{i\varepsilon t} U_3 \ket{\psi_T} U_3^{\dag} \ket{\psi_R} \right) \notag \\
    & + & \frac{1}{2} \ket{1} \left( \ket{\psi_T} \ket{\psi_R} - e^{i\varepsilon t} U_3 \ket{\psi_T} U_3^{\dag} \ket{\psi_R} \right) .
\end{eqnarray}
Thus, the probability of the $\ket{0}$ state in the ancilla qubit reflects the energy difference between the two states:
\begin{equation}
    P_0 = \frac{1}{2} \sum_{i<j} |\alpha_i|^2 |\beta_j|^2 \left( 1 + \cos(\Delta E_{ij} - \varepsilon) t \right) ,
\end{equation}
where $\Delta E_{ij} = \braket{u_i|\cH^z|u_i} - \braket{u_j|\cH^z|u_j}$. We assume that the test state is nearly in the ground state, and the reference state is a superposition of the ground state with probability $1-\delta$ and the first excited state with probability $\delta (\ll 1)$ when the $y$-field is applied in the correct direction. In this case, we can approximate the probability of the two-level system as follows:
\begin{eqnarray}
    P_0(\varepsilon) & = & \frac{1}{2} \left\{ 2 (1-\delta) + \delta(1+\cos(\Delta E - \varepsilon)t \right\} \\
    \label{eq12}
    & \sim & 1 - \delta \left( \frac{(\Delta E - \varepsilon) t}{2} \right)^2 ,
\end{eqnarray}
where $\Delta E$ is the energy gap between the ground state and the first excited state of the problem Hamiltonian. In the case where the $y$-field is applied in the incorrect direction, we assume the opposite condition, where the reference state is nearly in the ground state and the test state is in the superposition. This condition results in a change of sign, $\Delta E \to -\Delta E$, in Eq.~(\ref{eq12}).

The curve of $P_0(\varepsilon)$ forms an upward convex parabola. As $|\Delta E|$ increases, $P_0(0)$ decreases. Therefore, the variable (qubit) with the smallest $P_0(0)$ value is selected as the most sensitive one. This process corresponds to line 5 in Algorithm~\ref{alg1}. Once the most sensitive variable is identified, the sign of its sensitivity is determined by evaluating $P_0(\varepsilon) - P_0(0)$, as $P_0(\varepsilon)$ shifts the convex and breaks the symmetry. This step corresponds to line 6 in Algorithm~\ref{alg1}. If this value is negative, it implies that $\Delta E$ is also negative, indicating that the $y$-field decreases the energy and the variable can be fixed in the direction determined by the rotation induced by the $y$-field. On the other hand, if $P_0(\varepsilon) - P_0(0)$ is positive, the variable can be fixed in the opposite direction. By repeating this procedure to fix one sensitive variable at a time, all variables are eventually fixed to specific directions.

\subsection{DAQGO2: QA with control}

\begin{figure}[h]
    \centering
    \leavevmode
    \Qcircuit @C=1em @R=.7em {
      \lstick{\ket{0}} & \gate{R_x(\frac{\pi}{2})} & \ctrlo{1} & \ctrlo{1} & \ctrl{1}  & \ctrl{1} & \gate{Z} & \ctrl{2} & \gate{H} & \meter \\
      \lstick{\ket{0}} & \gate{H} & \ctrlo{1} & \ctrl{1} & \ctrlo{1}  & \ctrl{1} & \ctrl{-1} & \qw & \gate{H} & \meter \\
      \lstick{\ket{0}} & {/^N} \qw & \gate{U_1} & \gate{U_2} & \gate{U_1} & \gate{U_2} & \qw & \gate{U_3^\dag} & \qw & \qw
    }
    \caption{Quantum circuit for DAQGO2 and DAQGO3. $U_1$ and $U_2$ are controlled QA blocks used for the preparation of states with and without the $y$-field.}
    \label{fig2}
\end{figure}
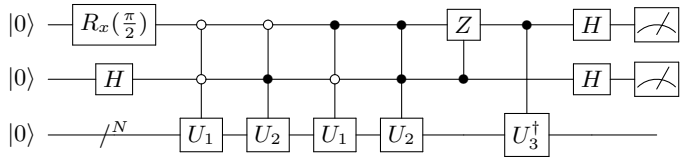

DAQGO1 is designed to minimize the size and complexity of the quantum circuit. However, this assumption is not suitable for short-time annealing scenarios where the ground state is not dominant in the final state. To address this, we have designed another circuit, DAQGO2, which can calculate the energy difference $\braket{\psi_T|\cH^z|\psi_T} - \braket{\psi_R|\cH^z|\psi_R}$ more accurately. DAQGO2 consists of four controlled quantum annealing blocks and employs two ancilla qubits for measurement and branch selection, as shown in Figure~\ref{fig2}. The rotation gate $R_x(\pi/2)$ in the circuit is expressed as
\begin{equation}
    R_x\left(\frac{\pi}{2}\right) = \frac{1}{\sqrt{2}} \begin{pmatrix} 1 & -i \\ -i & 1 \end{pmatrix} .
\end{equation}
Contrary to DAQGO1, DAQGO2 has controlled unitary evolution of $U_1$ and $U_2$, which are challenging to implement. Current QA devices do not possess such functionality.

The final state of the circuit is expressed as follows:
\begin{eqnarray}
    \ket{\psi}_F
    & = & \ket{00} \left( (\ket{\psi_T} + \ket{\psi_R}) - iU_3^\dag (\ket{\psi_T} - \ket{\psi_R})\right) / 4 \notag \\
    & + & \ket{01} \left( (\ket{\psi_T} - \ket{\psi_R}) - iU_3^\dag (\ket{\psi_T} + \ket{\psi_R})\right) / 4 \notag \\
    & + & \ket{10} \left( (\ket{\psi_T} + \ket{\psi_R}) + iU_3^\dag (\ket{\psi_T} - \ket{\psi_R})\right) / 4 \notag \\
    & + & \ket{11} \left( (\ket{\psi_T} - \ket{\psi_R}) + iU_3^\dag (\ket{\psi_T} + \ket{\psi_R})\right) / 4 . \notag \\
\end{eqnarray}
By selecting the $\ket{0}$ branch of the second qubit, the state is simplified as:
\begin{eqnarray}
    \ket{\psi}_{F}^{(0)}
    & = & \ket{00} \left( (\ket{\psi_T} + \ket{\psi_R}) - iU_3^\dag (\ket{\psi_T} - \ket{\psi_R})\right) / 2\sqrt{2} \notag \\
    & + & \ket{10} \left( (\ket{\psi_T} + \ket{\psi_R}) + iU_3^\dag (\ket{\psi_T} - \ket{\psi_R})\right) / 2\sqrt{2} . \notag \\
\end{eqnarray}
The probability difference between the $\ket{0}$ and $\ket{1}$ states of the first ancilla qubit is:
\begin{eqnarray}
    Q^{(0)} & = & P_0^{(0)} - P_1^{(0)} \notag \\
    & = & \frac{1}{2} \Im \left( (\bra{\psi_T} + \bra{\psi_R}) U_3^\dag (\ket{\psi_T} - \ket{\psi_R}) \right) .
\end{eqnarray}
For short durations of the controlled Hamiltonian simulation, $t \ll 1$, we can approximate the dynamics as $U_3^\dag = e^{i\cH^z t} \sim 1 + i\cH^z t$. Thus, the probability difference becomes:
\begin{eqnarray}
    Q^{(0)}(t) & \sim & \frac{t}{2}(\braket{\psi_T|\cH^z|\psi_T} - \braket{\psi_R|\cH^z|\psi_R}) \notag \\
    \label{eq17}
    & & - \Im \braket{\psi_T|\psi_R} .
\end{eqnarray}
If we select the $\ket{1}$ branch, $Q^{(1)}(t)$ has the same expression as $Q^{(0)}(t)$ except for the sign of the last term. Moreover, both branches are selected with equal probability. Thus, the energy difference can be calculated as:
\begin{equation}
g_i = (Q^{(0)}(t) + Q^{(1)}(t))/t .
\end{equation}
Therefore, Algorithm~\ref{alg1} can be implemented using DAQGO2.

\subsection{DAQGO3 and 4: imaginary part of inner product}

We also investigate the performance of other measures, which involve the imaginary part of the inner product $\Im \braket{\psi_T|\psi_R}$. Two variations of these measures exist: $Q^{(0)}(t)$ and $Q^{(0)}(0)$. As indicated in Eq.~(\ref{eq17}), the former measure includes an additional term reflecting the energy difference. DAQGO1 and DAQGO2 quantify the energy difference of states for the problem Hamiltonian, which is a weighted sum of energies of classical states. DAQGO3 and DAQGO4 utilize the imaginary part of the inner product of the two states, separately evolved by the total Hamiltonian, with and without the $y$-field, respectively.

\begin{figure}[h]
    \centering
    \leavevmode
    \Qcircuit @C=1em @R=.7em {
      \lstick{\ket{0}} & \gate{R_x(\frac{\pi}{2})} & \ctrlo{1} & \ctrl{1}  & \gate{H} & \meter \\
      \lstick{\ket{0}} & {/^N} \qw & \gate{U_2} & \gate{U_1} & \qw & \qw
    }
    \caption{Quantum circuit for DAQGO4.}
    \label{fig3}
\end{figure}
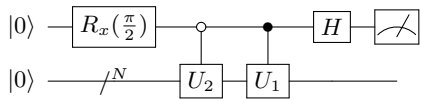

The phase difference information of the states encompasses the developmental paths of the two systems. Even if observables, such as energy, of QA for two states are identical, their phases might differ. By incorporating two states into a single circuit, we can measure the difference in phases by calculating the imaginary part of the inner product. The circuit for DAQGO4 can be simplified as depicted in Fig.~\ref{fig3}.

\subsection{Shot number estimation and experiments on a quantum processor}

To conduct experiments on a real device, we estimate the required shot number using sample size determination in statistics. In the sensitivity analysis, since energy measures vary, the statistical conditions differ. This requires individual determination of shot numbers for each case. Our sensitivity analysis consists of two steps: variable selection and sign determination. In the first step, sensitivities for all unfixed variables are evaluated and compared. To streamline this process, we estimate the sample size needed to distinguish between the best and the second-best variables. Then, we estimate the sample size for sign determination.

In QGO, we measure energy, so the estimation is conducted based on the normal distribution. In DAQGOs, measurement on a qubit follows a binomial distribution. For all estimations, the significance level is set to 0.99. Detailed information is provided in Appendix \ref{apxA}. Additionally, to reduce the number of shots, we discretize the parameter space of the algorithm ($\tau$, $t$, $\varepsilon$, and $|c_{\mathrm{opt}}|$) into a grid and perform a greedy search for optimization. Detailed dependencies of the parameters on performance are discussed in~\cite{Matsumori2023}.

Using the estimated shot numbers for DAQGO1, we conducted experiments on a 5-qubit circuit using the IonQ Harmony for $N = 2$ systems. The DAQGO1 algorithm employs QA for state preparation followed by a gate circuit. Currently, there is no hardware platform that combines QA and a gate circuit. Therefore, we prepare states using a quantum circuit~\cite{Mottonen2005} to emulate QA. The output states from QA are calculated beforehand using QuTiP~\cite{Johansson2013} and decomposed into CNOT gates and single-qubit rotation gates.

\section{Results}
\label{sec3}

\subsection{Algorithm comparison on simulator}

\begin{figure}[h]
    \centering
    \begin{overpic}[]{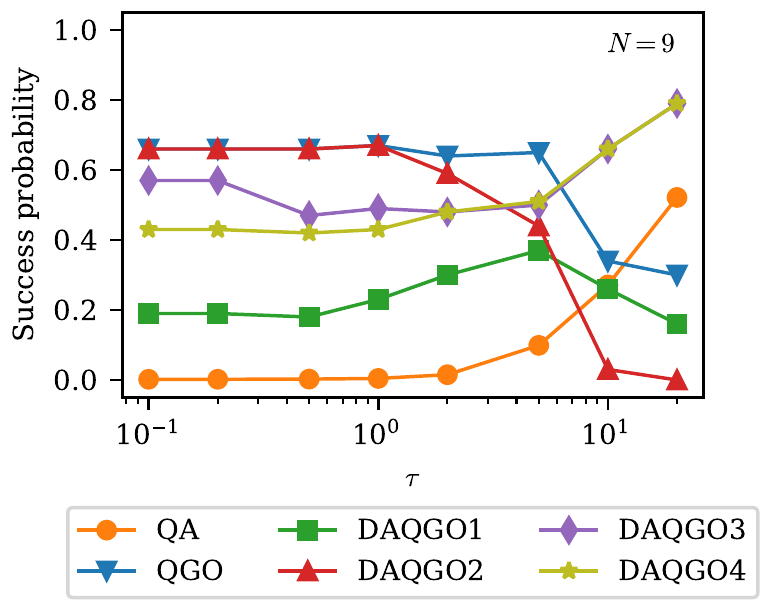}
        \put(0,79){(a)}
    \end{overpic} \\
    \vspace{0.2in}
    \begin{overpic}[]{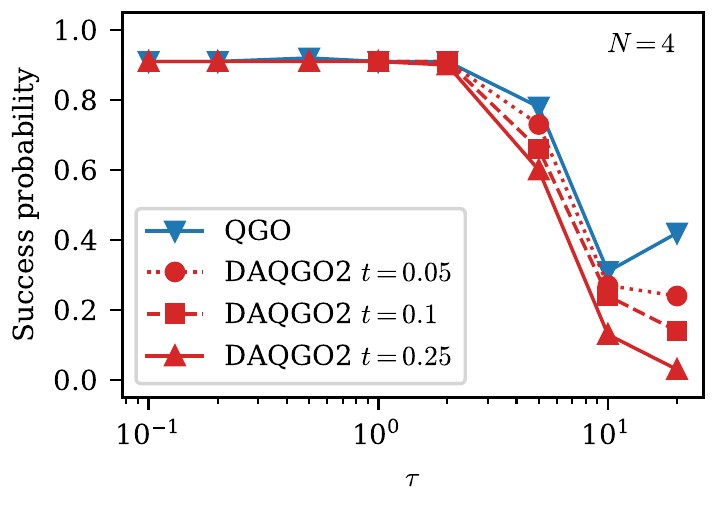}
        \put(0,70){(b)}
    \end{overpic}
    \caption{
        \label{fig4}
        Success probabilities plotted as functions of the annealing time $\tau$. (a) Performance comparison of various algorithms for an $N = 9$ system. (b) Performance for an $N = 4$ system across different Hamiltonian dynamics times.
    }
\end{figure}

Fig.~\ref{fig4}(a) presents the success probabilities among different algorithms and annealing times $\tau$ for a system size of $N=9$ in 100 random instances. The interaction and longitudinal field coefficients, $J_{ij}$ and $h_i$, were generated from a normal distribution with zero mean and a standard deviation of $1/(N-1)$. The choice of $N-1$ in the denominator is intended to adjust the energy level per spin for comparability across systems of different sizes~\cite{Kadowaki2023}. The numerical analyses for this and the following subsections were conducted using {\it mesolve} in QuTiP for analog circuits, followed by the operator-level circuit simulator in QuTip for digital circuits. The controlled unitary operations in the analog circuits, specifically controlled-$U_1$ and $U_2$, were formulated using either Eq.~\ref{eq6} or Eq.~\ref{eq7}. The final state of QA is a superposition of the ground and excited states. Therefore, to calculate the success probability for a given instance, we sum the probabilities of all degenerate ground states. Additionally, we average these probabilities across all instances. In contrast, the QGO and DAQGO algorithms output a specific solution. Therefore, the success probability is defined as the ratio of the number of successful instances to the total number of instances.

QGO and DAQGOs show improved performance in short annealing times because the sensitivity analysis identifies the preferable direction of each qubit in terms of energy or another measures, and then amplifies the qubit's probability by applying an optimal CD field strength to turn it in the preferable direction. However, for long annealing times, two main challenges arise: the test and reference states become similar, complicating the sensitivity analysis, and the CD term is an approximation. Consequently, there is no advantage if the conventional QA operates near adiabatic conditions.

The figure indicates that DAQGO1 is not an effective algorithm across a broad range of annealing times. In short annealing times, QGO and DAQGO2 exhibit better performance, indicating that accurate energy estimation is beneficial. Their results diverge as the annealing time $\tau$ increases. The energy calculation in DAQGO2 is approximated based on a short-time Hamiltonian simulation.

Fig.~\ref{fig4}(b) illustrates that the approximation improves with shorter simulation times. DAQGO3 and DAQGO4 consistently outperform QA. Moreover, they show superior performance to QGO in longer annealing times. DAQGO3 utilizes both the energy and the imaginary part of the inner product. The energy term contributes in short annealing times, but there is no significant difference in long annealing times as the performances of DAQGO3 and DAQGO4 are the same.

\begin{figure}[h]
    \centering
    \begin{overpic}[]{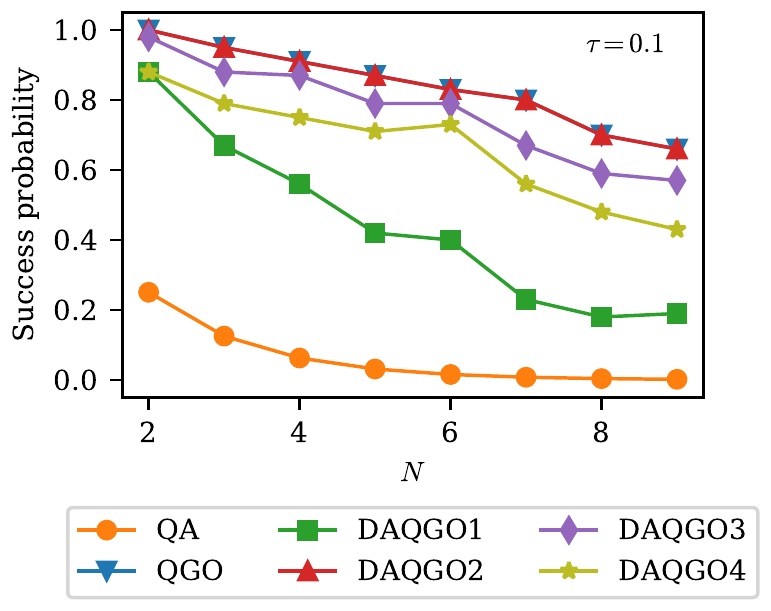}
        \put(0,79){(a)}
    \end{overpic} \\
    \vspace{0.2in}
    \begin{overpic}[]{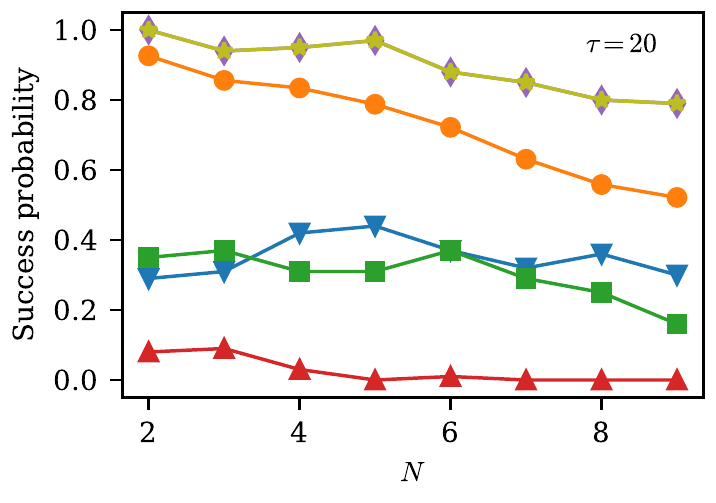}
        \put(0,70){(b)}
    \end{overpic}
    \caption{
        \label{fig5}
        Success probabilities plotted for different algorithms as functions of the system size $N$. (a) Performance for short annealing time, $\tau = 0.1$. (b) Performance for long annealing time, $\tau = 20$.
    }
\end{figure}

Figure~\ref{fig5} displays the dependency of success probability on system size for both short ($\tau = 0.1$) and long ($\tau = 20$) annealing times. Increase in the system size leads to a decrease in the energy gap between the ground and first excited states. This smaller energy gap facilitates transitions to the excited states, generally resulting in diminished performance with an increase in system size. Among the various algorithms tested under these two annealing times, DAQGO3 and DAQGO4 consistently exhibit robust performance irrespective of the annealing time and sample size.

\subsection{Shot number estimation}

\begin{figure}[h]
    \centering
    \begin{overpic}[]{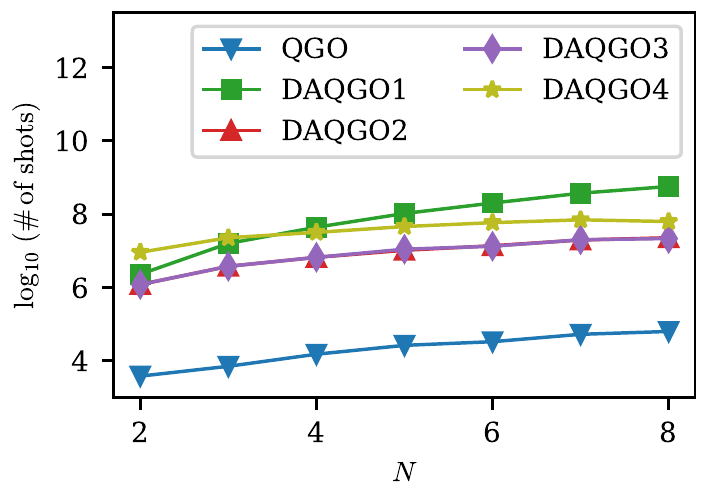}
        \put(0,70){(a)}
    \end{overpic} \\
    \vspace{0.1in}
    \begin{overpic}[]{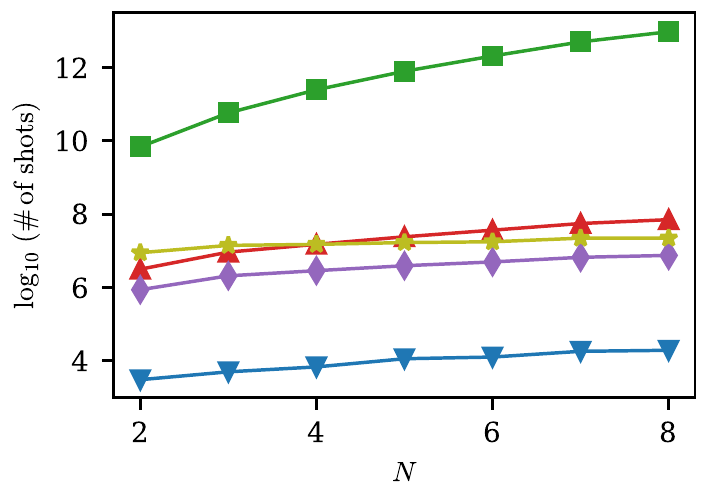}
        \put(0,70){(b)}
    \end{overpic}
    \caption{
        \label{fig6}
        Shot number estimations for different algorithms plotted as functions of the system size $N$. (a) Performance for variable selection. (b) Performance for sign determination.
        }
\end{figure}

Fig.~\ref{fig6} presents the estimated shot numbers for both variable selection and sign determination across all algorithms. The method for calculating the total number of shots per experiment varies by algorithm: In QGO, the energies of both reference and test states are measured for $|g_i|$ (variable selection) and $\sgn g_k$ (sign determination). To estimate the total number of shots, the larger value from Fig.\ref{fig6}(a) or (b) is chosen and multiplied by $N(N+3)/2$, representing the total count across all iterations. On the other hand, in DAQGO algorithms, energy differences are measured directly. In DAQGO1, which incurs significant costs in sign determination, the total number of shots is estimated by adding $N(N+1)/2$ times the shots for variable selection to $2N$ times the shots for sign determination. For DAQGO2-4, the estimation method is the same as in QGO, except the multiplication factor is $N(N+1)/2$. Among them, the original QGO requires the fewest shots. In energy estimation, QGO measures all qubits and maps the measurements to the energy. Conversely, DAQGO algorithms map the energy differences to an ancilla qubit. These differences are reflected in the shot number requirements. DAQGO1 displays lower performance and requires a larger number of shots. DAQGO2, 3, and 4 are comparable, but DAQGO3 requires slightly fewer shots.

\subsection{Experiments on the quantum processor}

\begin{figure}[h]
    \centering
    \begin{overpic}[]{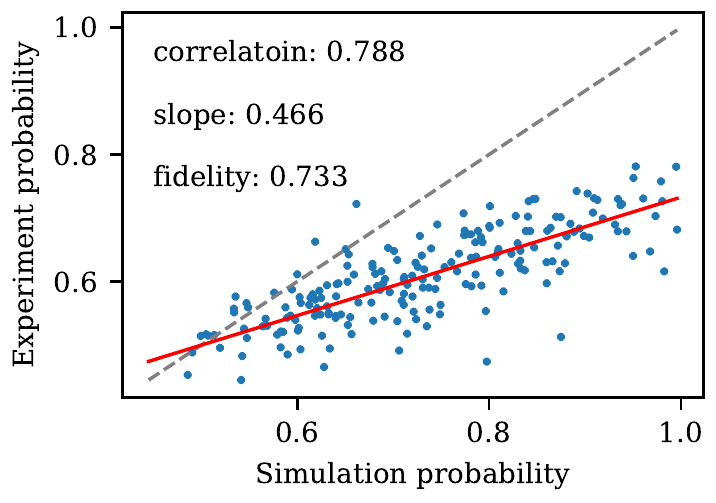}
        \put(0,70){(a)}
    \end{overpic} \\
    \vspace{0.2in}
    \begin{overpic}[]{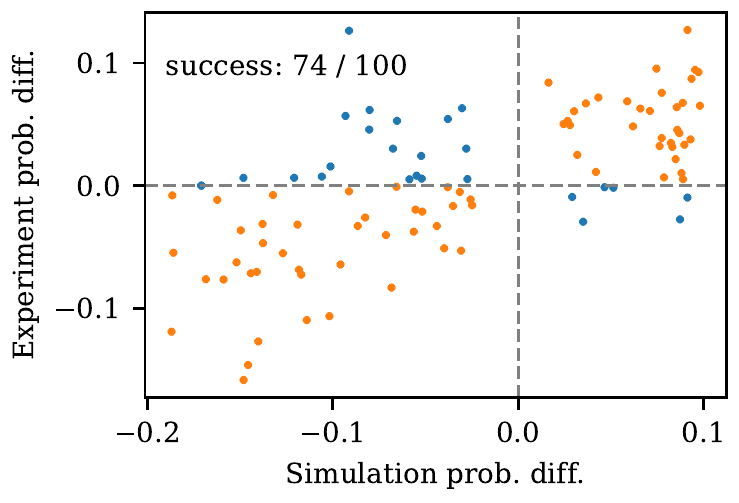}
        \put(0,70){(b)}
    \end{overpic}
    \caption{
        \label{fig7}
        (a) Comparison of experimentally measured probabilities and simulated probabilities, with a linear regression passing through the point (0.5, 0.5) indicated in red. (b) Comparison of estimated and simulated values used to determine the sign of the sensitivity. Data points located in the first and third quadrants represent successful results (marked in orange), while the others represent failed results (marked in blue).
    }
\end{figure}

Fig.~\ref{fig7}(a) compares the expected and observed probabilities. Optimized parameters for these experiments are $\tau = 0.6$, $t = 0.5$, $\varepsilon = 1.2$, and $|c_{\mathrm{opt}}| = 1.5$. We observe a linear relationship between the two probabilities of the ancilla qubit. At an expected probability of 1.0, the observed probability is 0.733. This discrepancy is due to device noise, and this value is the estimated fidelity of the circuit. Using the single-qubit and two-qubit operation fidelities (0.9974 and 0.9898 as of Oct. 26, 2022, respectively) provided by IonQ, the predicted fidelity is 0.744, where the circuit has 19 single-qubit and 24 two-qubit operations. The quantum circuit for this experiment is described in Appendix \ref{apxB}. The estimated and predicted values align well. Shot noise contributes to the deviation of the points from the red regression line.

Fig.~\ref{fig7}(b) plots the expected and observed values of $P_0(\varepsilon) - P_0(0)$. Points in the first and third quadrants correspond to successful results, while the others indicate failure. The success probability is 0.74, which is below the target value of 0.99 we set in the shot number estimation. This discrepancy is because the estimation assumes a fidelity of 1, i.e., we only consider shot noise. Using the actual fidelity and the shot number from our experiment, we can re-predict the success probability as 0.733. The predicted and actual success probabilities are also in good agreement.

\section{Discussion}
\label{sec4}

In this work, we have explored algorithms within the DAQC framework to expedite the realization of quantum computing applications through effective allocation of digital and analog circuits. We have proposed Quantum Annealing (QA) within the DAQC framework (DAQGO), which incorporates either (1) QA as a state preparation or (2) a controlled QA block.

The first algorithm, DAQGO1, employs a straightforward digital circuit to approximate energy calculation, but it does not demonstrate improvement over the original QGO. The algorithms in the second category (DAQGO2-4) have deeper circuits than the first. DAQGO2 calculates energy in the digital circuit and shows performance comparable to QGO. The measures in DAQGO3 and DAQGO4 include $\Im \braket{\psi_T|\psi_R}$, which cannot be calculated efficiently through classical means by measuring these states and subsequently constructing the inner product. Their performances surpass both QGO and QA in longer annealing times.

The primary challenge with the second category of algorithms lies in the difficulty of implementing controlled QAs. Recent studies demonstrate that shallow, gate-based CD circuits enhance their performance~\cite{Hegade2021,Hegade2022,Chandarana2022,Guan2023,Matsumori2023}. This advancement potentially enables the digitized implementation of controlled QAs. Nonetheless, the efficiency of such implementations remains an open question, as this area of research is still in its nascent stages.

DAQGO3 and DAQGO4 exemplify that the quantum processing of quantum data could have advantages over the classical processing of quantum data. More specifically, phase information, which reflects the trajectory of the dynamics and is unique to quantum mechanics, is utilized to improve performance in our proposed algorithms. In other words, while QGO measures QA outputs individually, DAQGOs perform quantum gate operations among states derived from different QA conditions with and without $y$-field. Similar advantage of quantum processing of quantum data in machine learning is demonstrated~\cite{Huang2022}.

We have employed statistical sample size determination methods to estimate the required number of shots. This estimation aids in comparing algorithms for specific problem instances, going beyond standard order analysis or comparisons with fixed shot numbers.

We have not conducted time-to-solution (TTS) analysis, which measures the time needed to find the ground state with a given probability, such as 0.99. Although QGO exhibits an advantage in success probability over QA, it's comparable in terms of TTS~\cite{Kadowaki2023}. To achieve a shorter TTS, both a higher success probability and a fewer shot number are required. Designing an algorithm that takes both factors into account is an important consideration. However, this question is not addressed in this paper and is left for future work.

For experiments on a quantum processor, we show that the results of DAQGO1 with $N=2$ align closely with theoretical predictions. However, given the current fidelity of the processor, systems with $N > 2$ and DAQGO2-4 algorithms are infeasible. For instance, DAQGO1 with $N=3$ requires about 160 circuit depth, resulting in a fidelity under 0.5. Therefore, further advancements in both digital and analog quantum devices are required. In particular, a QA system with a control qubit and arbitrary input state could lessen the burden on the digital component. Additionally, improving the algorithm's implementation for existing devices is another promising direction for future work.

Furthermore, in the early era of fault-tolerant quantum computing, future work could consider error correction or mitigation strategies~\cite{Pudenz2014,Suzuki2022} for the analog part of DAQGO and other integrations of QA within the DAQC framework.

\begin{acknowledgments}
The author would like to express his gratitude to Hidetoshi Nishimori, Yuichiro Matsuzaki, and Takashi Imoto for their insightful discussions. A part of this work was performed for Council for Science, Technology and Innovation (CSTI), Cross-ministerial Strategic Innovation Promotion Program (SIP), “Promoting the application of advanced quantum technology platforms to social issues”(Funding agency: QST).
\end{acknowledgments}

\appendix

\section{Sample size determination}
\label{apxA}

In the QGO algorithm, we employ the means and variances of the energy distributions of two QA outputs. We assume the sample size (the shot number) is large enough to perform z-test. The sample size can be determined based on the definition of the z-score,
\begin{equation}
    z = \frac{d}{\sqrt{\frac{\sigma_1^2}{N} + \frac{\sigma_2^2}{N}}} ,
    \label{eqA1}
\end{equation}
where $z$ is the z-score, $d$ is the energy difference to be detected, $\sigma_1^2$ and $\sigma_2^2$ are the variances of the energy distributions of the two QA outputs, and $N$ is the sample size. The z-score must exceed the critical z value ($z^*$) associated with a specified significance level, for example, 0.99,
\begin{equation}
    z \ge z^* .
    \label{eqA2}
\end{equation}
By substituting Eq.~\ref{eqA1} into Eq.~\ref{eqA2}, we calculate the estimated sample size as
\begin{equation}
    \label{eqA3}
    N \ge \frac{{z^*}^2 (\sigma_1^2 + \sigma_2^2)}{d^2} .
\end{equation} 

\begin{figure}[h]
    \centering
    \includegraphics[]{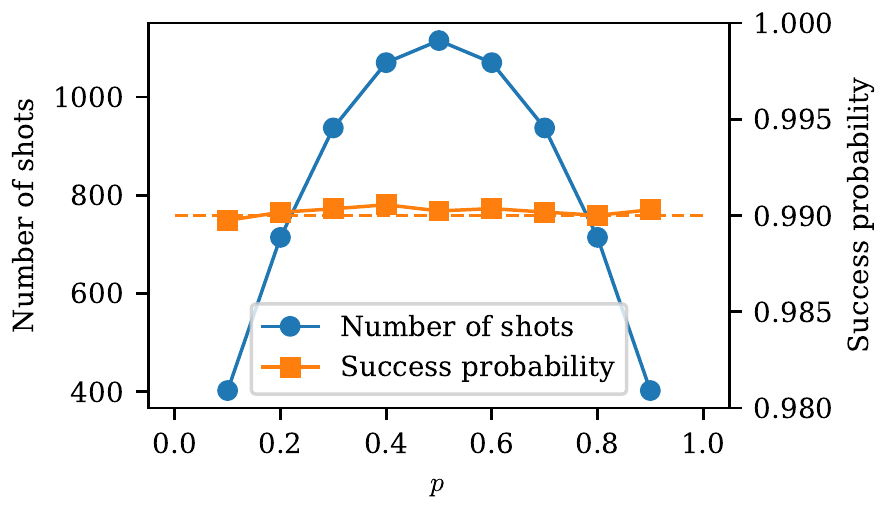}
    \caption{
        \label{fig8}
        The sample size (number of shots) is determined using the Wald method. The success probability refers to the ratio of successful tests to total tests, where the data sets are sampled from a Bernoulli distribution of a given probability.
        }
\end{figure}

In DAQGOs, energy is measured through an ancilla qubit. Hence, the measurements follow a binomial distribution. In this scenario, the sample size can be determined by employing the Wald method. By applying the z-score within this context,
\begin{equation}
    z = \frac{d}{\sqrt{\frac{p(1-p)}{N}}} ,
\end{equation}
the estimated sample size is given as
\begin{equation}
    N \ge \frac{{z^*}^2 p (1-p)}{d^2} \times \sqrt{2},
\end{equation}
where $p$ is the probability of the ancilla qubit, and $d$ is the probability difference between two groups. We multiply by a factor of $\sqrt{2}$ to more accurately determine the sample size in two-sample testing. A numerical demonstration is shown in Fig.~\ref{fig8}. In this analysis, the significance level was set to 0.99 (corresponding to the critical z value of 2.58), and $d$ was set to 0.05. For each calculated sample size $N$, we generated 100,000 data sets with probabilities of $p - d/2$ and $p + d/2$. We tested whether the data set could accurately distinguish the difference between the two groups. As depicted in the figure, about 99\% of sampled data sets correctly distinguished the two groups.

\section{DAQGO1 circuit implementation}
\label{apxB}

\begin{turnpage}

\begin{figure}[p]
    \centering
    \includegraphics[width=24cm]{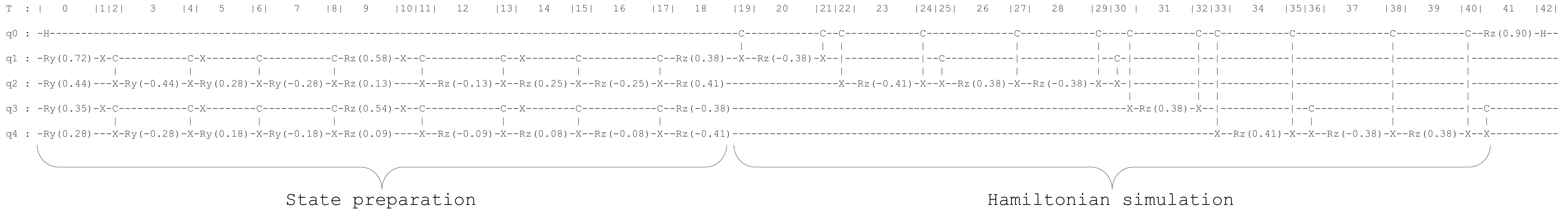}
    \caption{
        \label{fig9}
        A quantum circuit for DAQGO1 experiments on the quantum processor. The ancilla qubit is designated as q0. In the circuit, two quantum systems with $N=2$ are represented as q1-q2 (test system) and q3-q4 (reference system). The CNOT gate is indicated by ``C'' for the control qubit and ``X'' for the target qubit. Rotation angles depend on the specific details of each instance. 
        }
\end{figure}
\end{turnpage}

Fig.~\ref{fig9} shows the quantum circuit for the DAQGO1 algorithm applied to an optimization problem with $N=2$. To mimic the outputs of two QAs, $U_1$ and $U_2$, general state preparation circuits are used, as directly simulating the QA processes efficiently in a gate-based circuit is not feasible. This mimicry is achieved by pre-calculating the final QA outputs using a QA simulator implemented in QuTiP.

\clearpage
\clearpage

\bibliography{main}

\end{document}